\documentclass[preprint,12pt,3p]{elsarticle}
\usepackage{lineno,hyperref}
\usepackage{amsmath}
\usepackage{graphicx}
\usepackage{epstopdf}
\usepackage{float}
\usepackage{multirow}
\usepackage{hhline}
\usepackage{xcolor}
\usepackage{subcaption}
\usepackage{lineno}

\usepackage{setspace}

\makeatletter
\newcommand{\vast}{\bBigg@{4}}
\newcommand{\Vast}{\bBigg@{5}}
\def\ps@pprintTitle{%
	\let\@oddhead\@empty
	\let\@evenhead\@empty
	\let\@oddfoot\@empty
	\let\@evenfoot\@oddfoot
}
\makeatother

\begin{document}
\begin{frontmatter}

\title{Low Energy Light Yield of Fast Plastic Scintillators}

\author[1]{T.A. Laplace\corref{cor1}}
\ead{lapthi@berkeley.edu}
\cortext[cor1]{Corresponding author}

\author[1]{B.L. Goldblum}
\author[1,2]{J.A. Brown}
\author[3]{D.L. Bleuel}
\author[1,3]{C.A. Brand}
\author[1]{G. Gabella}
\author[1]{T. Jordan}
\author[1]{C. Moore}
\author[1]{N. Munshi}
\author[1]{Z.W. Sweger}
\author[1]{A. Ureche}
\author[2]{E. Brubaker}

\address[1]{Department of Nuclear Engineering, University of California, Berkeley, California 94720 USA}
\address[2]{Sandia National Laboratories, Livermore, California 94550, USA}
\address[3]{Lawrence Livermore National Laboratory, Livermore, California 94550 USA}

\begin{abstract}
Compact neutron imagers using double-scatter kinematic reconstruction are being designed for localization and characterization of special nuclear material. These neutron imaging systems rely on scintillators with a rapid prompt temporal response as the detection medium. As n-p elastic scattering is the primary mechanism for light generation by fast neutron interactions in organic scintillators, proton light yield data are needed for accurate assessment of scintillator performance. The proton light yield of a series of commercial fast plastic organic scintillators---EJ-200, EJ-204, and EJ-208---was measured via a double time-of-flight technique at the 88-Inch Cyclotron at Lawrence Berkeley National Laboratory. Using a tunable deuteron breakup neutron source, target scintillators housed in a dual photomultiplier tube configuration, and an array of pulse-shape-discriminating observation scintillators, the fast plastic scintillator light yield was measured over a broad and continuous energy range down to proton recoil energies of approximately 50~keV. This work provides key input to event reconstruction algorithms required for utilization of these materials in emerging neutron imaging modalities.   
\end{abstract}


\end{frontmatter}

\section{Motivation}

The ability to detect and image special nuclear material is important for nuclear security applications. Current neutron imaging systems are bulky and their sensitivity is limited by their geometric efficiency \cite{vanier, mascarenhas, marleau, bravar}. The development of a compact system capable of high-efficiency fast neutron imaging can enable weak threat source detection on a human-portable platform, where directional information can be used for background discrimination in low signal-to-noise environments. Advanced neutron imagers, including single-volume scatter camera (SVSC) concepts where two neutron scattering events are detected within a single volume, are being designed at Sandia National Laboratories to support these goals~\cite{braverman, weinfurther}. The work presented here addresses the measurement of the energy-dependent proton light yield of candidate scintillator materials for use in these novel neutron imagers. 

The proton light yield, or the relationship between light output and proton energy deposited in an organic scintillator, varies non-linearly with recoil energy due to quenching phenomena \cite{birks}. For neutron scatter camera imaging modalities, knowledge of the proton light yield is necessary to convert the measured light into proton recoil energy, which is then used along with n-p elastic scattering kinematics to determine the energy and direction of the incident neutron. As both the magnitude of the scintillation intensity and the quenching factor are properties of the scintillating medium, the proton light yield is a characteristic unique to each scintillator material formulation. Existing physics-based models of proton light yield give an unreliable extrapolation to recoil energies below 1~MeV \cite{norsworthy} and it is in this low energy region where the prompt fission neutron spectrum of special nuclear material peaks. Eljen Technology's EJ-200, EJ-204, and EJ-208 scintillators are potential candidates for the SVSC due to their fast rise times and relatively long ($>1.5$~m) optical attenuation lengths. In this work, the proton light yield of these fast plastic scintillators was extracted in the energy range of approximately 50~keV up to 5~MeV using a double time-of-flight (TOF) technique. 

\section{Methods}

\begin{figure}
\center
\includegraphics[width=0.8\textwidth]{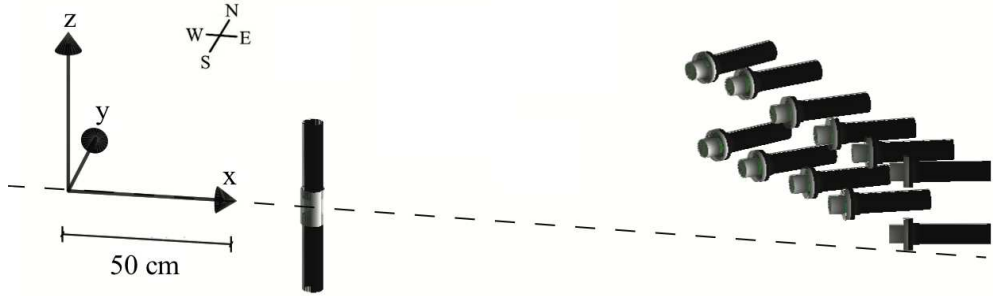}
\caption{Experimental setup for proton light yield measurements. The neutron beam traveled along the $x$-axis (illustrated with a dashed line) through the dual-PMT target scintillator cell. Eleven observation detectors were positioned at forward angles with respect to the incoming neutron beam. \label{exp-setup}}
\end{figure}

Neutrons were produced by impinging a 16-MeV $^2$H$^+$ beam from the 88-Inch Cyclotron at Lawrence Berkeley National Laboratory onto a 3-mm-thick Be breakup target located in the cyclotron vault. This deuteron-breakup neutron source provides an intense, broad spectrum neutron beam for experimental inquiry~\cite{covo}. The target scintillator to be characterized (i.e., EJ-200, EJ-204, or EJ-208) was placed in beam, as shown in Figure~\ref{exp-setup}. Out of beam and at different forward angles, eleven EJ-309 pulse-shape-discriminating organic liquid scintillators from Eljen Technology were positioned to serve as observation detectors in the coincident measurement of scattered neutrons from n-p elastic scattering events in the target cell. 

Using the method of Brown et al.\ \cite{brown-DTOF, brown-thesis}, the neutron TOF between the breakup target and the target scintillator was used to determine the energy of the incident neutron, $E_n$. Likewise, the neutron TOF between the target scintillator and the observation detectors was used to determine the energy of the scattered neutron, $E_n^{\prime}$. The proton recoil energy, $E_p$, was then calculated using the incoming neutron energy and the known neutron scattering angle, $\theta$:
\begin{linenomath*}
\begin{equation}
\label{thisone}
E_p = E_n \sin^2\!{\theta}. 
\end{equation}
\end{linenomath*}
The proton recoil energy associated with a given target scintillator pulse was determined on an event-by-event basis to provide a continuous measurement of the proton light yield relation. The outgoing TOF was used for disambiguation of the incident beam pulses in cases where fast neutrons from a given cyclotron pulse had the same observable TOF as slower neutrons from previous pulses \cite{harrig}. This approach also provides strong rejection criteria for multiple scatters in the active target volume. 

\subsection{Experimental Configuration}

The experimental setup was configured to target proton recoil energies down to tens of keV. The observation detectors were placed such that the lowest angles were approximately $10^\circ$ with respect to the axis defined by incoming beam. Table~\ref{locationsDec} provides a measurement of the detector locations for the experiment. The origin of the coordinate system is defined at the beamline center along the west wall of the experimental area. The $x$-coordinate was taken as the distance from the west wall, the $y$-coordinate as the distance north from the beamline center, and the $z$-coordinate as the distance above and below the beamline center. The flight path between the Be breakup target and the origin of the coordinate system was $646.9~\pm~0.5$~cm along the $x$-axis, resulting in incident neutron flight paths to the target scintillators of approximately 7~m. The flight path between the target scintillator and the observation detectors ranged from $1.4$~m to $1.9$~m. 

\begin{table*} 
\centering
	\renewcommand{\arraystretch}{1.2}
	\begin{tabular}{ccccc}
		\hline
		Material    & ID &		x (cm)			&	y (cm)			&	z (cm)		\\ \hline
		EJ-200	& - 		&		52.4	$\pm$	0.5& $	0.0	\pm	0.5	$   &$	\textcolor{white}{-}0.0	\pm	0.5	$\\ 
		EJ-204     & - 		&   52.8	$\pm$	0.5& $	0.0	\pm	0.5	$   &$	-0.5	\pm	0.5	$\\ 
		EJ-208     & - 		&  52.0	$\pm$	0.5& $	0.0	\pm	0.5	$   &$	-0.2	\pm	0.5	$\\ 
		EJ-309	& 0		&	146.2	$\pm$	1	& $	104.0	\pm	1	$&$	\textcolor{white}{-}10.8	\pm	0.2	$\\ 
		EJ-309	& 1 		&	150.8	$\pm$	1	&$	98.2	\pm	1	$&$	-11.1	\pm	0.2	$\\ 
		EJ-309	& 2 		&	160.2 $\pm$	    1	&$	89.8	\pm	1	$&$	\textcolor{white}{-}11.9	\pm	0.2	$\\ 
		EJ-309	& 3 		&	165.0	$\pm$	1	&$	83.5	\pm	1	$&$	-11.2	\pm	0.2	$\\ 
		EJ-309	& 4		&	175.7	$\pm$	1	&$	74.7	\pm	1	$&$	\textcolor{white}{-}9.6	\pm	0.2	$\\ 
		EJ-309	& 5		&	182.6 $\pm$	    1	&$	66.9	\pm	1	$&$	-9.1	\pm	0.2	$\\ 
		EJ-309	& 6 		&	191.1 $\pm$	    1	&$	59.0	\pm	1	$&$	\textcolor{white}{-}9.7	\pm	0.2	$\\ 
		EJ-309	& 7		&	196.6 $\pm$	    1	&$	51.7	\pm	1	$&$	-9.2	\pm	0.2	$\\ 
		EJ-309	& 8		&	204.5 $\pm$	    1	&$	45.3	\pm	1	$&$	\textcolor{white}{-}9.8	\pm	0.2	$\\ 
		EJ-309	& 9 		&	217.9 $\pm$	    1	&$	31.7	\pm	1	$&$	\textcolor{white}{-}9.8	\pm	0.2	$\\ 
		EJ-309	& 10		&	218.1 $\pm$	    1	&$	30.1	\pm	1	$&$	-9.2\pm	0.2	$\\ 
		\hline
	\end{tabular}
	\caption{Summary of detector materials, identification (ID) numbers, locations, and estimated uncertainties in the measurements. \label{locationsDec}}
\end{table*}

The three target detectors were composed of 5.08 cm dia.\ $\times$ 5.08 cm len.\ right circular cylindrical scintillator cells mounted in a dual-photomultiplier tube (PMT) configuration. The cylinder sides were wrapped with several layers of polytetrafluoroethylene tape to reflect the scintillation light and each circular base was coupled to a Hamamatsu H1949-51 PMT, except for one base of the EJ-208 cell, which was coupled to a Hamamatsu H1949-50 PMT. The eleven observation detectors were composed of 5.08 cm dia.\ $\times$ 5.08 cm len.\ right circular cylindrical cells of EJ-309 contained in a thin aluminum housing and mounted to Hamamatsu H1949-50 and H1949-51 PMTs on one end of the cylindrical cells. The optical coupling between the scintillator cells and the borosilicate glass window of the PMTs was achieved using BC-630 silicone grease. The PMTs for the target and observation detectors were powered using CAEN NDT1470 and CAEN~R1470ETD high voltage power supplies, respectively. The observation detector gains were set to accommodate pulse heights corresponding roughly to the maximum proton recoil energy anticipated for a given scattering angle.  

To accept small signals but reject background from thermionic emission of the photocathodes, each target scintillator signal was taken by accepting only coincident pulses from the two PMTs. This, along with a signal in one of the observation detectors, was required to trigger the system. The RF control signal from the cyclotron was set to propagate when an event from both target PMTs and an observation cell occurred. The bias voltage of the PMTs in the dual-PMT target configuration was selected to roughly gain match the detectors by setting the full height of signals corresponding to the Compton edge of the $662$~keV $\gamma$~ray from a $^{137}$Cs source at 75\% of the 2~V full scale range of the digitizer. The target cell gain settings ranged from 1650~V to 2000~V. To provide a systematic check and cover a broader proton recoil energy range, the EJ-204 proton light yield data were measured using two independent bias voltage settings of the dual-PMT target.

Data were acquired over a period of approximately 10~h for each bias voltage setting of the target scintillators with a beam current of approximately 40~nA. The data were recorded using a CAEN V1730 500 MS/s digitizer. The arrival time of the scintillator signals was established using the CAEN digital constant fraction discrimination algorithm, with a 75$\%$ fraction and 4~ns delay \cite{DPP-PSD}. The timing pick-off for the cyclotron RF signal was accomplished using leading edge discrimination. Full waveforms of the events were written to disk in list mode with global time stamps for pulse processing and event reconstruction in post-processing.

\subsection{Signal Processing}

Data reduction was accomplished in an object-oriented C\texttt{++} framework, which used elements of the ROOT data analysis framework~\cite{brun}. Waveforms from the scintillator cells were reduced to pulse integrals with 200~ns integration lengths ensuring collection of $95\%$ of the scintillation light. The PMT response functions were measured for all target PMTs at each of the run-time bias voltages using a finite difference method adapted from Friend et al.\ \cite{friendLinearity} and the linearity of the responses was confirmed.

To finely balance the signals from the two target PMTs in post-processing, a $^{137}$Cs source was placed $>5$~cm from the target scintillator to ensure uniform distribution of the electron recoils in the $z$-dimension and pulse integral spectra were taken. The minimum $\chi^2$ method was used to adjust the pulse integral of one signal relative to the other assuming a linear scaling with a resolution term to allow for variability amongst the response of the two PMTs~\cite{dietze}. The signals from each PMT were then combined using a geometric average to obtain the total scintillation light independent of the interaction location~\cite{knoll}. The resulting signal was then calibrated as described in the following section. 

\subsection{Light Output Calibration}
\label{lightCalib}

Given the known non-proportionality of the electron response of EJ-200 \cite{nassalski,payne} and the similar formulation of EJ-200, EJ-204, and EJ-208, the electron-equivalent light unit is not a useful quantity for these materials. As such, the light output was determined relative to that of a 477~keV electron, as evaluated using the Compton edge of a 662~keV $\gamma$ ray. Using a $^{137}$Cs source placed $>5$~cm from the target scintillator, pulse integral spectra were obtained using the summed target signal output. Simulation of the energy deposition spectrum from a 662~keV $\gamma$ ray was also performed using Geant4 \cite{geant4} and convolved with a detector resolution function \cite{dietze}. The simulation geometry included the target scintillator cell, the thin aluminum sleeve, and the PMT borosilicate glass window and magnetic shield. The channel number corresponding to the Compton edge of the 662~keV $\gamma$ ray was obtained by minimizing $\chi^2$ between the measured and simulated source spectra. The parameter minimizations were performed using the SIMPLEX and MIGRAD algorithms from the ROOT Minuit2 package, and error matrices were obtained using the HESSE algorithm \cite{brun}. The relative light unit was then defined as the light corresponding to a given proton recoil energy relative to the light produced by a 477~keV recoil electron. 

\subsection{Timing Calibrations}

\begin{figure}
	\center
	\includegraphics[width=0.8\textwidth]{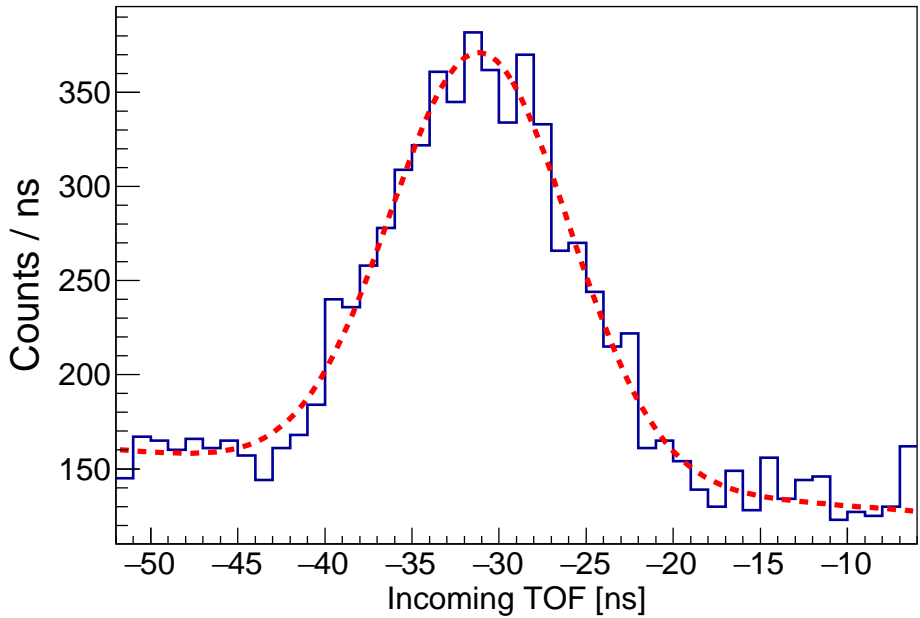}
	\caption{(Color online) The blue curve shows a histogram of time differences between the cyclotron RF signal and $\gamma$-ray events in the EJ-200 target detector. The width of the pulse is primarily reflective of the spatial spreading of the beam pulse, with $\sigma=5.1$~ns. The red dashed curve is a fit of the measured data with a normal distribution plus linear background term. \label{incTOF}}
\end{figure}

Timing calibrations were performed for both the incident and outgoing TOF measurements as described in Ref.~\cite{brown-DTOF}. 
The incoming TOF, $t_{inc}$, can be represented as:
\begin{linenomath*} 
\begin{equation}
\label{incTOFEq}
t_{inc} = \frac{t_1 + t_2}{2} - t_{\mathrm{RF}}-t_c+n\times T_{cyc},
\end{equation}
\end{linenomath*}
where $t_1$ and $t_2$ are the clock times of the event in the top and bottom PMTs of the dual-PMT target, respectively, $t_\mathrm{RF}$ is the clock time of the cyclotron RF signal, $t_c$ is a time calibration constant, $n$ is an integer offset corresponding to the different cyclotron pulses, and $T_{cyc}$ is the cyclotron pulse period. A coincidence window was applied in post-processing for the two target PMT signals, enforcing that events observed were within 10~ns of each other. To obtain the time calibration constant, a histogram of the arrival time of events in the target cell was constructed. A binned maximum likelihood estimation between the photon flash arising from deuteron breakup on the Be target and the superposition of a normal distribution with a linear background term was performed. The centroid of the photon of the flash was obtained and $t_c$ was then set using the speed of light and the known flight path. This is illustrated in Fig.~\ref{incTOF} for the EJ-200 target scintillator. The uncertainty in the incoming TOF determination is dominated by the uncertainty in $t_c$ with the spreading of the photon flash being $\sigma=5.1$~ns. The cyclotron pulse period, $T_{cyc}$, was 158.5~ns, resulting in an integer ambiguity, $n$, in the incoming flight time due to frame overlap, where slower neutrons from previous pulses have the same arrival time as fast neutrons from a given pulse. 

The timing calibration of the outgoing TOF was performed using $\gamma-\gamma$ coincidences between the dual-PMT target and observation detectors in a procedure similar to that described above. The outgoing TOF, $t_{out}$, was taken as: 
\begin{linenomath*}
\begin{equation}
t_{out} = t_3-\frac{t_1+t_2}{2}-t_c,
\end{equation}
\end{linenomath*}
where $t_3$ is the clock time of the event in the observation cell. The calibration time constant was determined by creating a histogram of the $\gamma-\gamma$ coincidences, with pulse shape discrimination applied in the observation detectors, and fitting using the procedure described above. To resolve the integer ambiguity, $n$, in the incoming TOF described in Eq.~\ref{incTOFEq}, an event-by-event comparison was made between the measured incoming TOF and expected incoming TOF as determined via kinematics using the scattering angle and outgoing TOF. The event reconstruction included a requirement that the absolute time difference between the expected and measured incoming TOF was less than 15.8~ns, or $<10$\% of the cyclotron period. 

\subsection{Data Reduction}

The relativistic energy-time relationship was employed to determine the incoming and outgoing neutron energy from the respective TOF. With the known scattering angle, $\theta$, the proton recoil energy was then determined on an event-by-event basis using Eq.~\ref{thisone}. The data were grouped into three sets based on the ID number of the observation detector (see Table~\ref{locationsDec}): ID 0-4, ID 5-8, and ID 9-10, covering an angle range of $31-48^\circ$, $17-27^\circ$, and $10-11^\circ$, respectively, with respect to the axis defined by the incoming beam.

\begin{figure*}[t!]
	\centering
	\begin{subfigure}{0.3\textwidth}
		\centering
		\includegraphics[width=0.97\textwidth]{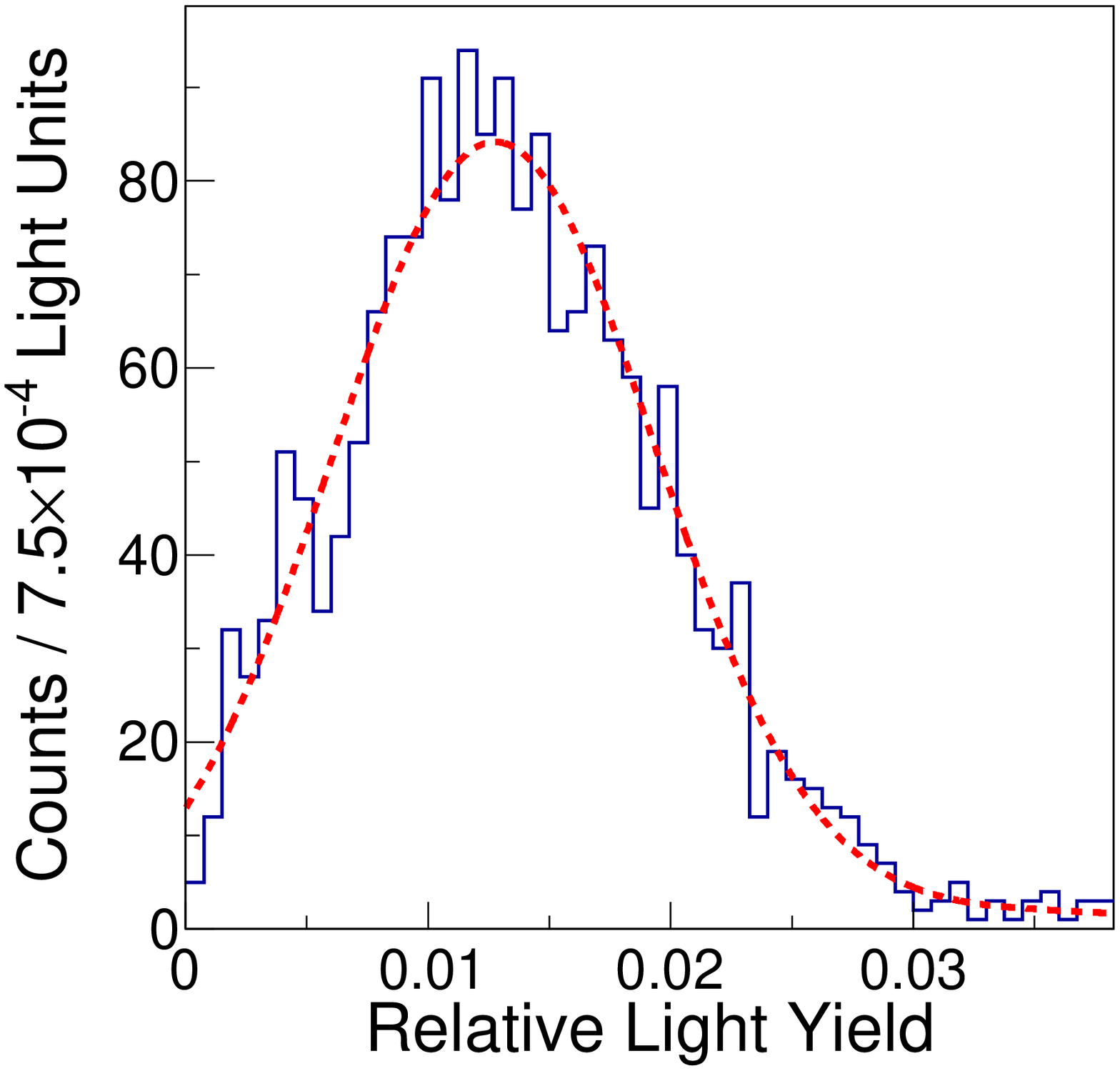}
		\caption{E$_p$ = 64 keV, ID $9-10$. \label{lowest}}
	\end{subfigure}%
	~ 
	\begin{subfigure}{0.3\textwidth}
		\centering
		\includegraphics[width=0.97\textwidth]{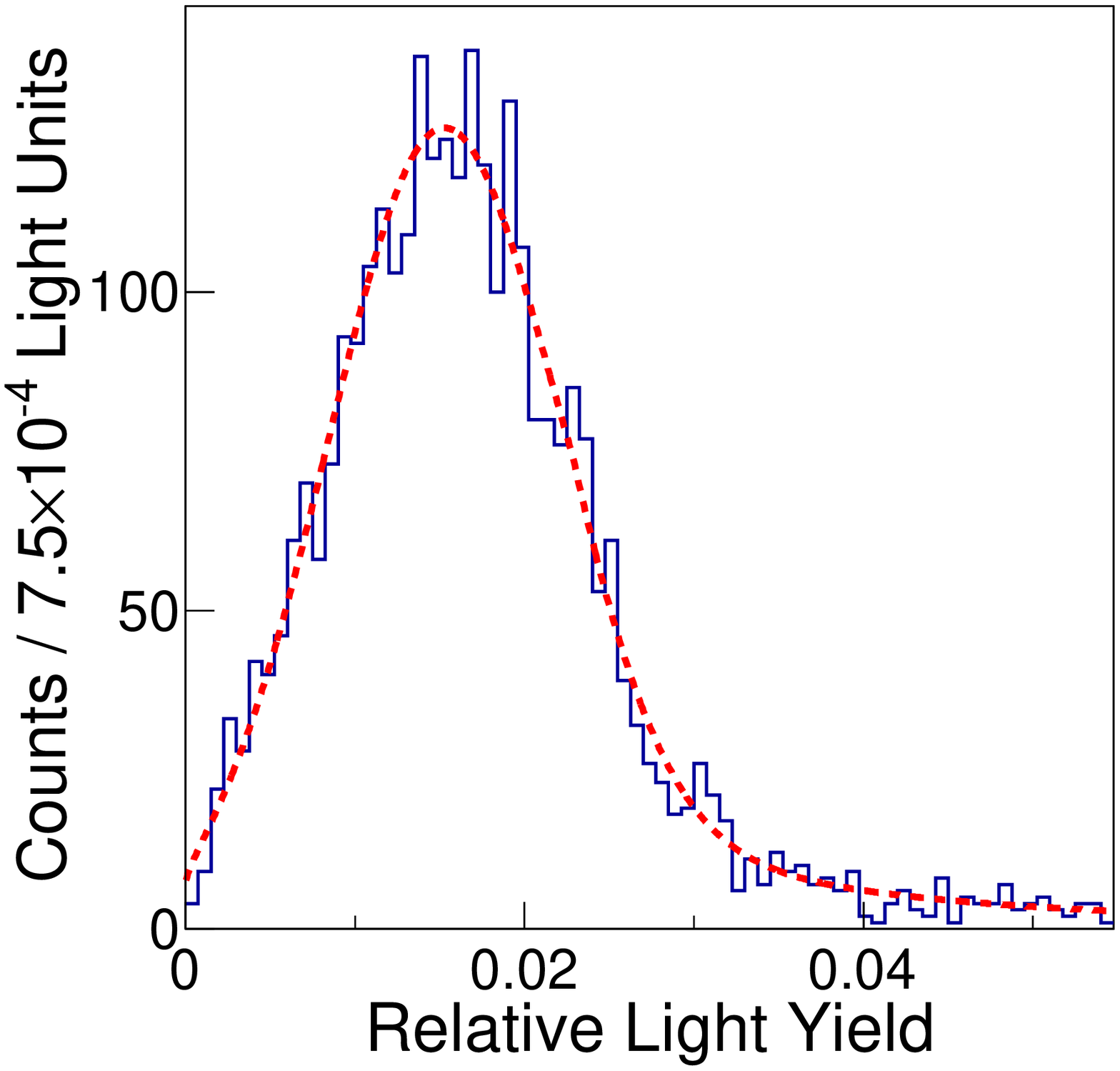}
		\caption{E$_p$ = 86 keV, ID $9-10$.}
	\end{subfigure}
    ~
    \begin{subfigure}{0.3\textwidth}
    	\centering
    	\includegraphics[width=0.97\textwidth]{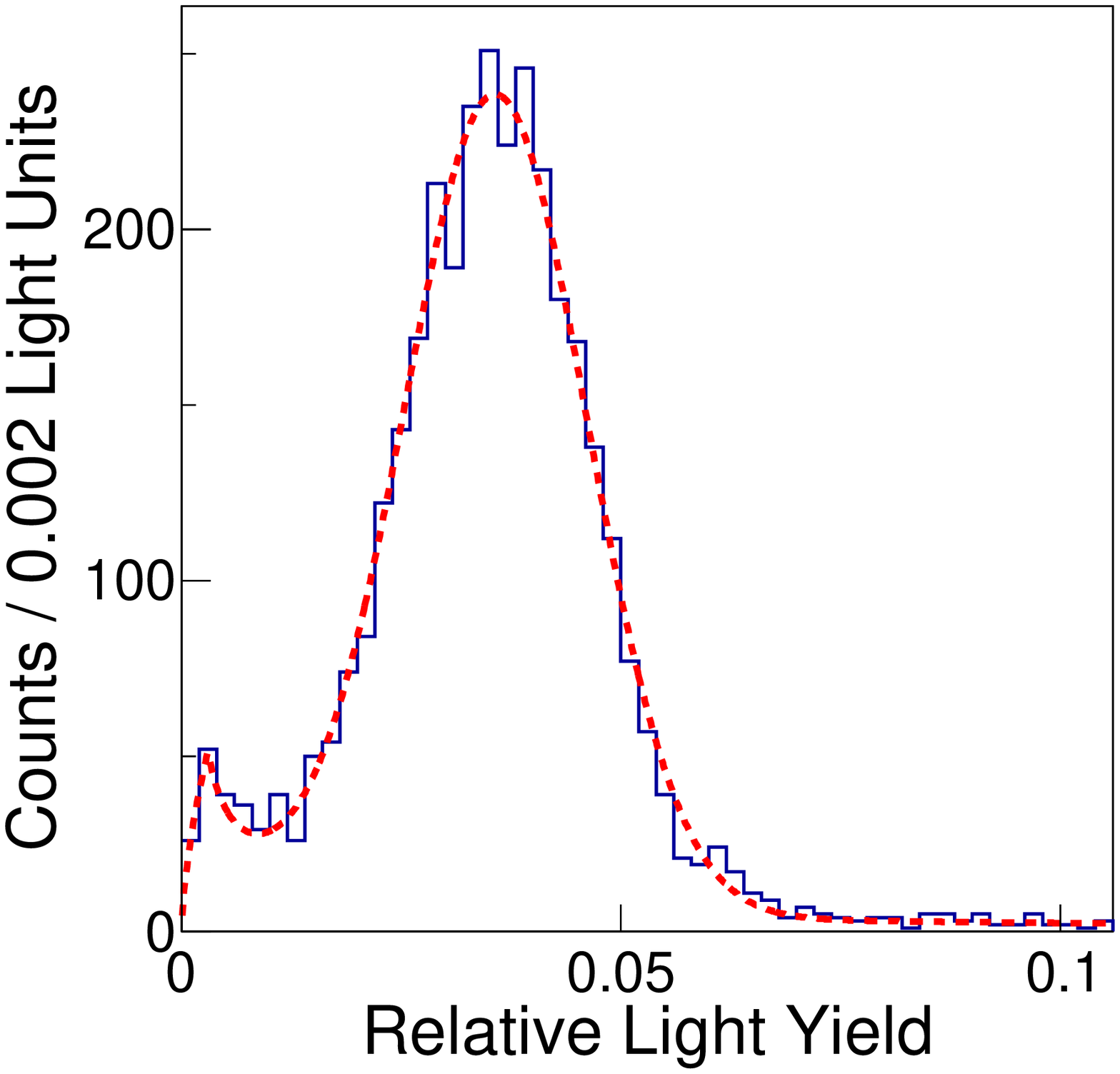}
    	\caption{E$_p$ = 202 keV, ID $5-8$.}
    \end{subfigure}
	
	\begin{subfigure}{0.3\textwidth}
		\centering
		\includegraphics[width=0.97\textwidth]{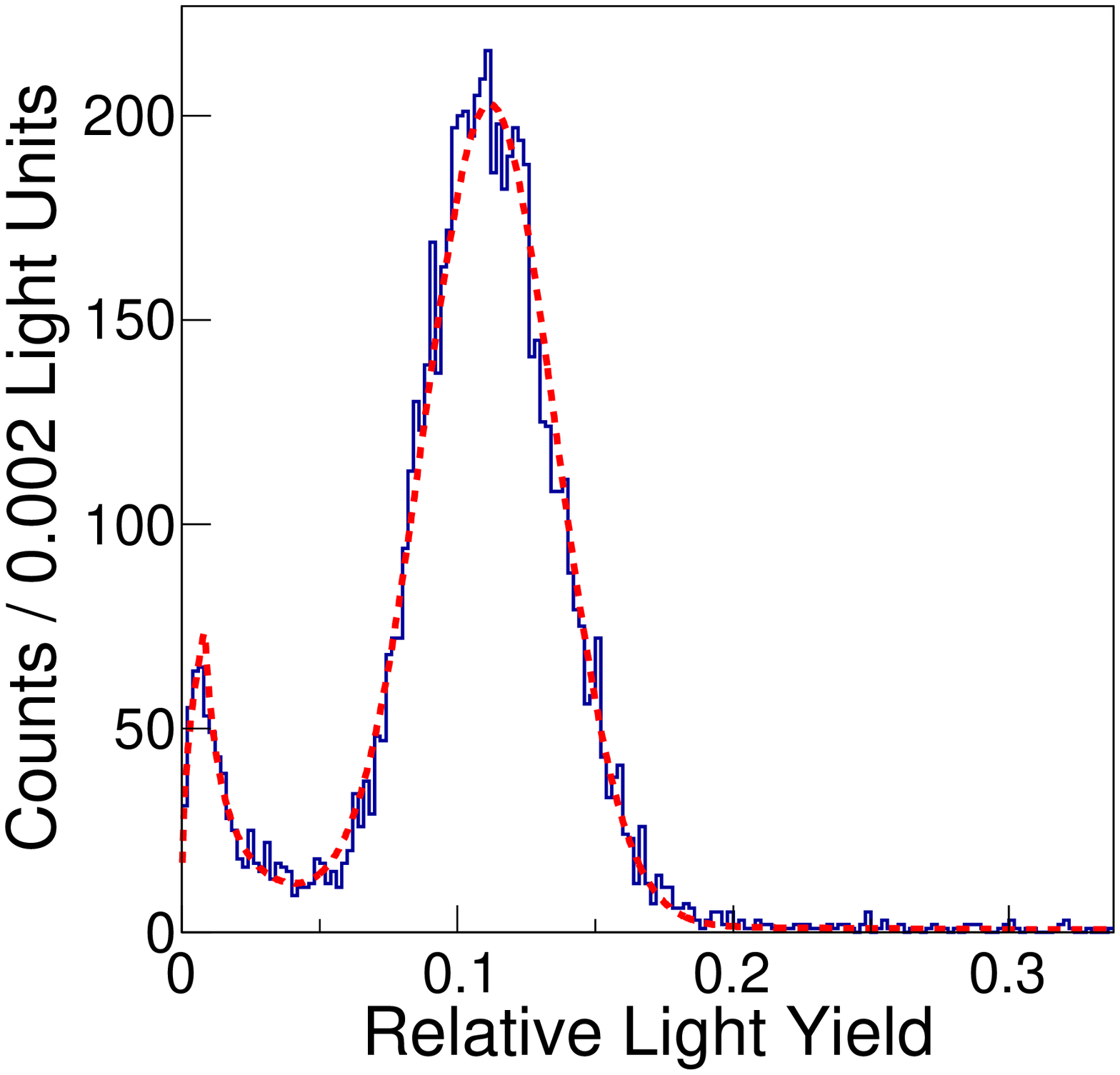}
		\caption{E$_p$ = 451 keV, ID $5-8$.}
	\end{subfigure}%
	~ 
	\begin{subfigure}{0.3\textwidth}
		\centering
		\includegraphics[width=0.97\textwidth]{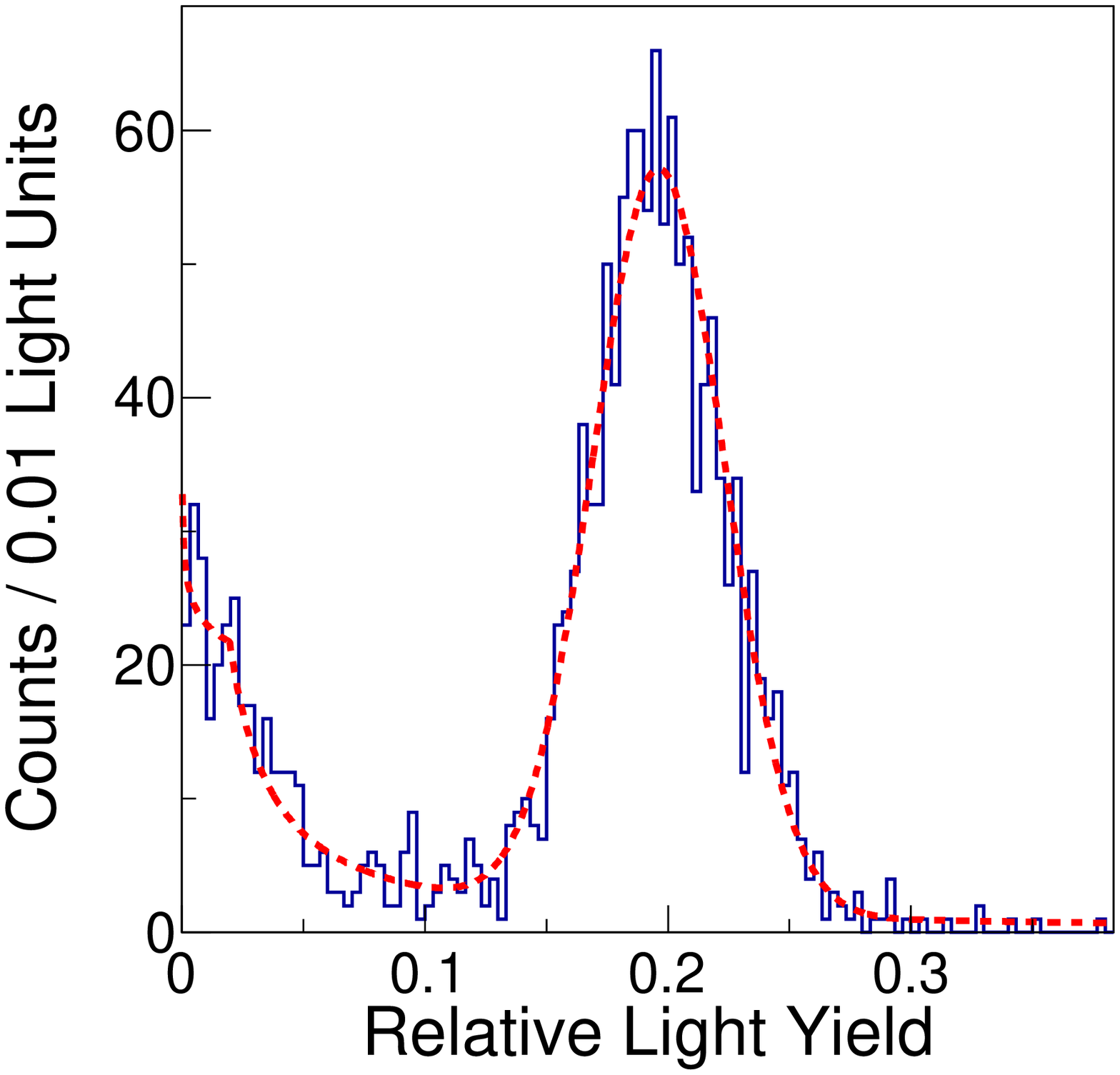}
		\caption{E$_p$ = 653 keV, ID $0-4$.}
	\end{subfigure}
	~ 
    \begin{subfigure}{0.3\textwidth}
	   \centering
	   \includegraphics[width=0.97\textwidth]{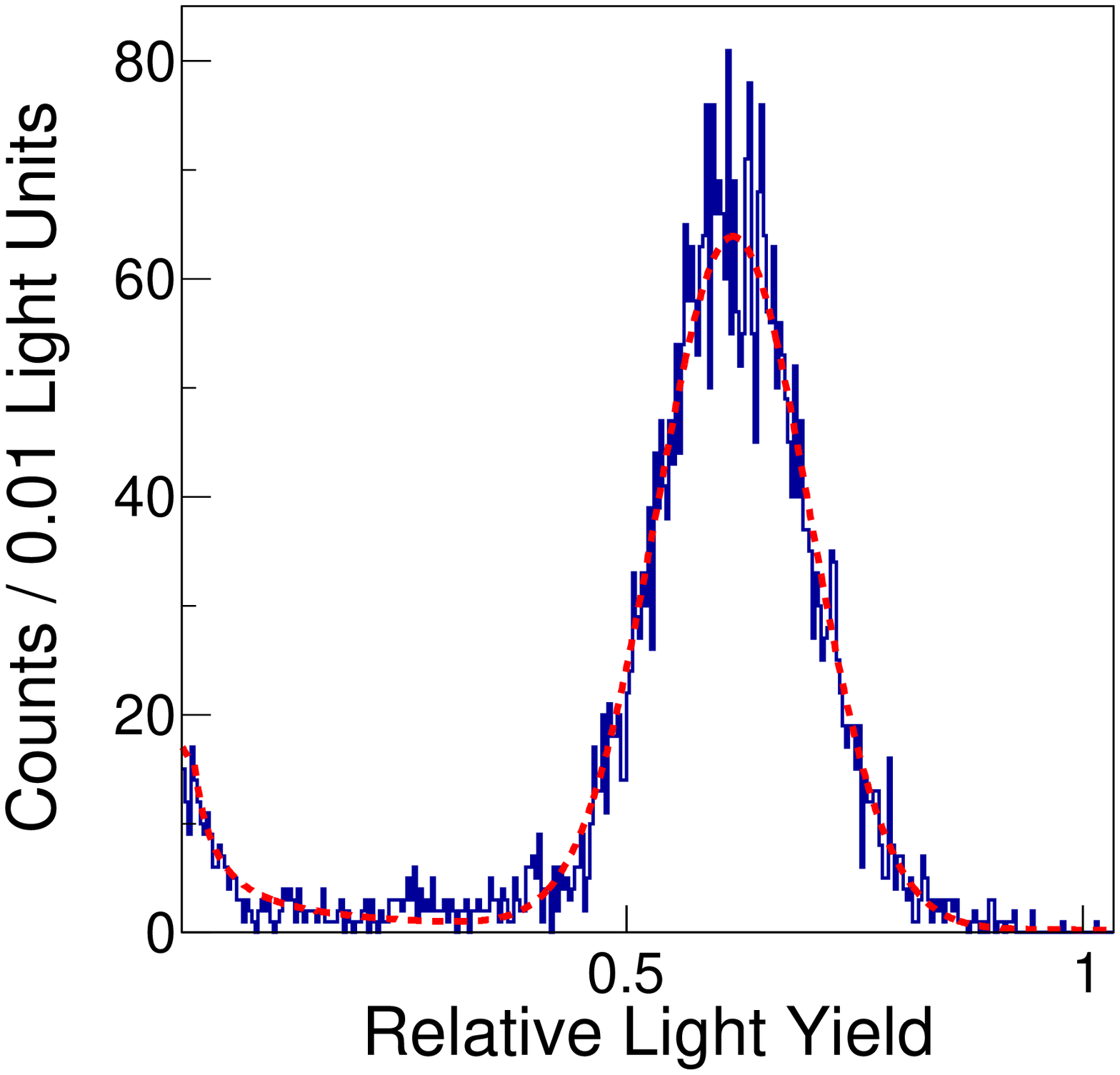}
	   \caption{E$_p$ = 1344 keV, ID $0-4$.}
    \end{subfigure}
	\caption{(Color online) Relative light yield spectra (blue curves) for a given proton energy bin fit with a piecewise power law and a normal distribution (red dashed curves) for the EJ-200 measurement. The centroid of the normal distribution corresponds to the mean light production for n-p elastic scattering events within the bin. The ID numbers specify the observation detectors used to observe this feature.}
	\label{sliceFit}
\end{figure*}

The data were reduced into a two-dimensional histogram of light output versus proton recoil energy, where the proton energy discretization corresponded to the lowest energy resolution for the detector group considered. The resolution was calculated using the uncertainty in the incoming TOF calibration, the uncertainty in the flight path, and an angular variance of 0.8$^\circ$, the latter obtained via a Monte Carlo simulation of the detector array. To help ensure events included in the histogram arose from single n-p elastic scattering events in the dual-PMT target, pulse shape discrimination was used to isolate neutron events in the observation cells and a maximum outgoing neutron energy constraint was applied reflective of the beam energy and known detector configuration. To produce a series of data points, projections of individual proton energy bins were made on the light yield axis. The resulting histograms were fit using a binned maximum likelihood estimation considering the peak to be normally distributed and the background to be represented by a continuous piecewise power law with two subdomains. Representative fits of the relative light yield spectra for proton recoil energies from 64~keV to 1.34~MeV for EJ-200 are shown in Fig.~\ref{sliceFit}. The coincident trigger of the dual-PMT target, the trigger detection threshold, and n-$\gamma$ rejection criteria in the observation detectors are potential sources of bias in the relative light distributions for low proton recoil energies.

\subsection{Uncertainty Analysis}
\label{UQ}
   
The sources of systematic uncertainty in the measurement of the light output corresponding to a given proton recoil energy include uncertainty in the determination of the incoming and coincident TOF temporal calibration constants, uncertainty in the detector locations, and uncertainty in the measured distance from the breakup target to the west wall of the experimental cave. To address the uncertainty from these potential sources of bias, a Monte Carlo method was implemented where the reduction from histogrammed quantities to data points was repeated while the input to calculations of the proton energy and light output were varied. The uncertainty in the channel number corresponding to the $^{137}$Cs edge determination was added in quadrature with the uncertainty obtained from the Monte Carlo simulation to provide the total uncertainty on the relative light yield.

\section{Results and Discussion}

\begin{figure}
	\center
	\includegraphics[width=0.8\textwidth]{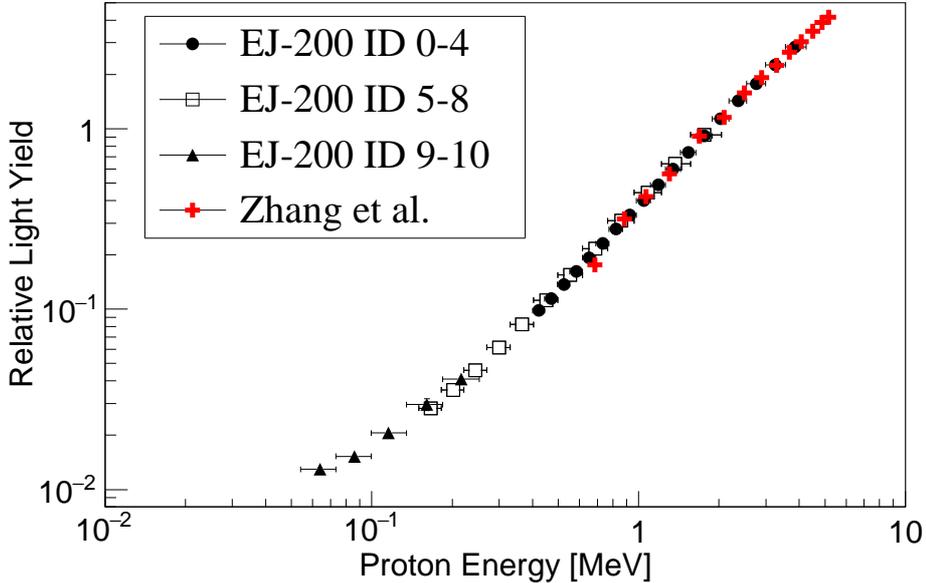}
	\caption{(Color online) Relative light yield of EJ-200 as a function of proton recoil energy. The ID numbers specify the observation detectors used to obtain the result in a given proton recoil energy range. \label{200LY}}
\end{figure}

Figure~\ref{200LY} shows a plot of the relative proton light yield of EJ-200. As described in Sec.~\ref{lightCalib}, the relative light unit is the light corresponding to a given proton recoil energy relative to the light produced by a 477~keV recoil electron. The data from this work are presented in three groups corresponding to three different neutron scattering angular ranges. The $x$-error bars reflect the proton energy resolution for the lowest scattering angle in a given observation detector grouping to provide a conservative estimate. The $y$-error bars include both the statistical and systematic uncertainty and are generally smaller than the data points. There exists a kinematic overlap, where multiple observation detectors view the same proton recoil energy range in the regions of approximately $150-250$~keV and $0.4-2$~MeV. The agreement in the results obtained from independent observation detectors provides additional confidence in the measurement. This work is also compared to a measurement from Zhang et al.\cite{zhang} of the proton light yield of BC-408, a commercial equivalent to EJ-200 from Saint-Gobain Crystals, obtained using an edge characterization technique. The measurements exhibit excellent agreement over the full energy range examined in this work.

\begin{figure}
	\center
	\includegraphics[width=0.8\textwidth]{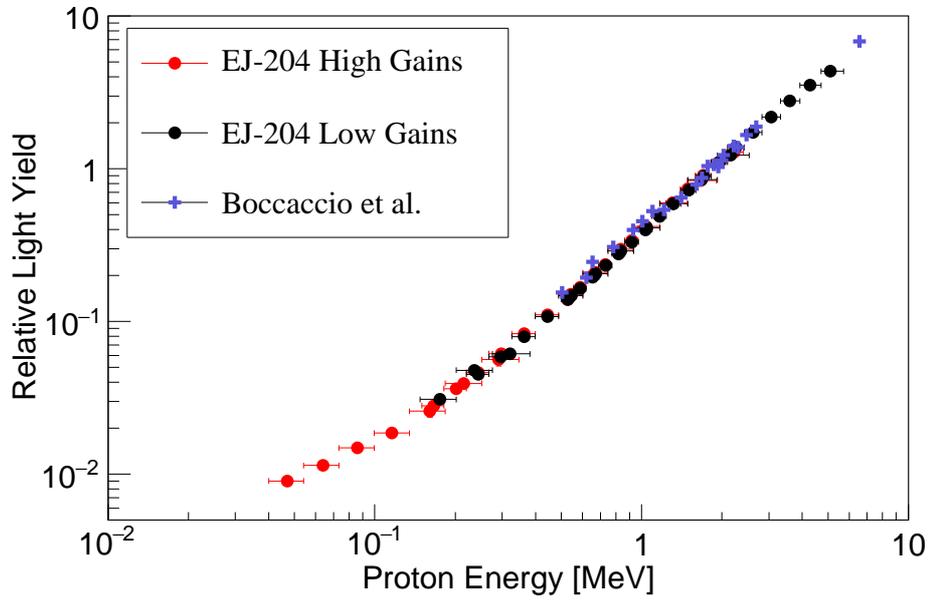}
	\caption{(Color online) Relative light yield of EJ-204 as a function of proton recoil energy. \label{204LY}}
\end{figure}

With a decay time of 1.8~ns (compared to 2.1~ns and 3.3~ns for EJ-200 and EJ-208, respectively), the EJ-204 plastic scintillator is attractive for faster timing experiments \cite{eljen}. Figure~\ref{204LY} shows a plot of the relative proton light yield of EJ-204. The EJ-204 scintillator light yield was measured for two different gain settings of the dual-PMT arrangement and excellent agreement was observed between the two measurements. The data for the high gains measurement extend down to 47~keV, providing a new low energy limit for proton light yield data on this material. A comparison to a measurement by Boccaccio et al.\ of a commercial equivalent, Nuclear Enterprise's NE-104, obtained using an edge characterization technique also demonstrates good agreement \cite{boccaccio}.

\begin{figure}
	\center
	\includegraphics[width=0.8\textwidth]{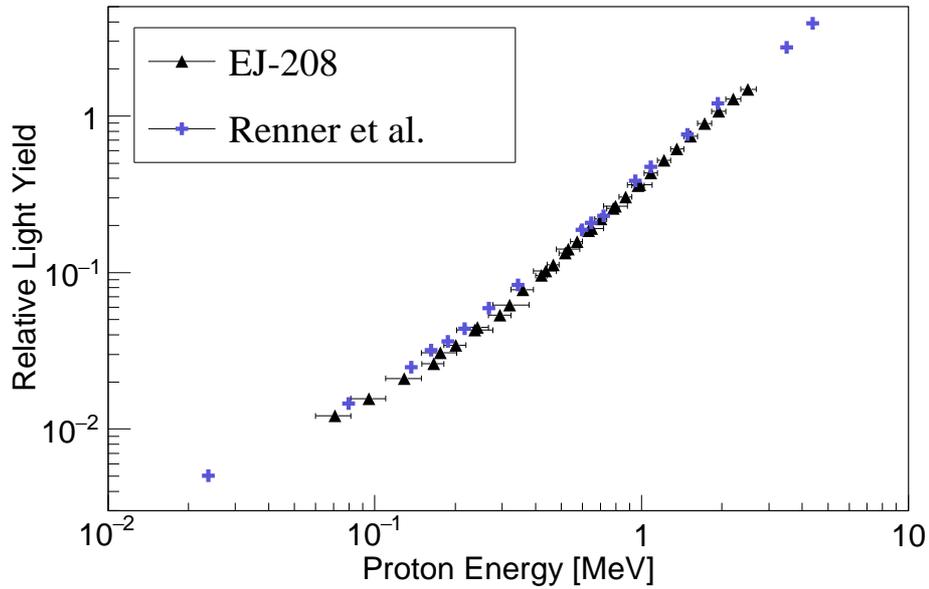}
	\caption{(Color online) Relative light yield of EJ-208 as a function of proton recoil energy. \label{208LY}}
\end{figure}

Figure~\ref{208LY} shows a plot of the relative proton light yield of EJ-208. The data were compared with a measurement from Renner et al.\ on a commercial equivalent, Nuclear Enterprise's NE-110, obtained using edge characterization without a Monte Carlo normalization~\cite{renner}. The results demonstrate general agreement over the full energy range examined within the estimated uncertainty, though the data from Renner et al.\ tend to lie systematically above the EJ-208 result. A summary table of the light yield data obtained in this work is provided in~\ref{appendix}.

\section{Summary}

The proton light yield of EJ-200, EJ-204, and EJ-208 was measured using a double TOF technique over a proton recoil energy range of approximately 50~keV to 5~MeV. These were compared to proton light yield data from commercial equivalents and a general consistency in the results was observed. Similar proton light yield relations were also observed for the three materials, which is reasonable given that they each employ the same polyvinyltoluene polymer base and have similar carbon and hydrogen atomic densities. The low energy proton light yield data obtained in this work enable the benchmarking of physics-based models of the specific luminescence of organic scintillators, which can in turn enhance understanding of organic scintillator physics and support the design of new detector materials. This work further supports the development of passive neutron imaging systems for nuclear security applications by enabling improved accuracy and neutron energy resolution in imagers employing double-scatter kinematic reconstruction. 

\section*{Acknowledgements}

The authors thank the 88-Inch Cyclotron operations and facilities staff for their help in performing these experiments. Thanks also to Walid Younes for expert advice and useful insights. This material is based upon work supported by the U.S. Department of Energy, National Nuclear Security Administration, Office of Defense Nuclear Nonproliferation Research and Development (DNN R\&D) through the Nuclear Science and Security Consortium under Award DE-NA0003180, Lawrence Berkeley National Laboratory under Contract DE-AC02-05CH11231, and Lawrence Livermore National Laboratory under Contract DE-AC52-07NA27344. Sandia National Laboratories is a multimission laboratory managed and operated by National Technology and Engineering Solutions of Sandia LLC, a wholly owned subsidiary of Honeywell International Inc., for the U.S. Department of Energy's National Nuclear Security Administration under contract DE-NA0003525. 

\appendix
\section{Light Yield Data}
\label{LYDataPoints}
\label{appendix}
The relative proton light yield data for EJ-200 and EJ-208 obtained using a 200~ns integration length are summarized in Table~\ref{resultsTable}. The relative proton light yield data for EJ-204 obtained using a 200~ns integration length for different gain settings of the dual-PMT target are summarized in Table~\ref{resultsTable204}. The asymmetric proton recoil energy error bars are reflective of the non-uniform distribution of proton energies within a given bin. 

\begin{table*}
	\centering
	\renewcommand{\arraystretch}{1.2}
	\setlength{\tabcolsep}{4pt}
	\begin{tabular}{ccc|ccc}
		\hline
		\multicolumn{3}{c|}{EJ-200} & \multicolumn{3}{c}{EJ-208}   \\
		Observation  & Proton Recoil   & Relative Light Yield  & Observation  & Proton recoil  & Relative Light Yield  \\
		       Detector ID     &  Energy [MeV] &  [dimensionless]   &   Detector ID & energy [MeV] & [dimensionless] \\
		\hline
		9-10 & 0.064$_{-0.010}^{+0.009}$ & 0.0129 $\pm$ 0.0006 &  9-10 & 0.071$_{-0.011}^{+0.010}$ & 0.0121 $\pm$ 0.0006 \\ 
		9-10 & 0.086$_{-0.013}^{+0.013}$ & 0.0153 $\pm$ 0.0010 &  9-10 & 0.095$_{-0.014}^{+0.015}$ & 0.0156 $\pm$ 0.0008 \\  
		9-10 & 0.115$_{-0.016}^{+0.020}$ & 0.0206 $\pm$ 0.0012 &  9-10 & 0.129$_{-0.019}^{+0.020}$ & 0.0211 $\pm$ 0.0017 \\ 
		9-10 & 0.161$_{-0.026}^{+0.023}$ & 0.0296 $\pm$ 0.0023 &  9-10 & 0.176$_{-0.027}^{+0.027}$ & 0.0307 $\pm$ 0.0019 \\ 
		9-10 & 0.216$_{-0.032}^{+0.037}$ & 0.0409 $\pm$ 0.0027 &  9-10 & 0.237$_{-0.034}^{+0.040}$ & 0.0428 $\pm$ 0.0027 \\ 
		&                           &                     &  9-10 & 0.320$_{-0.044}^{+0.058}$ & 0.0620 $\pm$ 0.0047 \\
		5-8 & 0.166$_{-0.016}^{+0.016}$ & 0.0282 $\pm$ 0.0011  &        &                       &                      \\
		5-8 & 0.202$_{-0.020}^{+0.020}$ & 0.0357 $\pm$ 0.0014  &  5-8 & 0.166$_{-0.016}^{+0.015}$ & 0.0261 $\pm$ 0.0016  \\
		5-8 & 0.245$_{-0.023}^{+0.025}$ & 0.0460 $\pm$ 0.0018  &  5-8 & 0.201$_{-0.020}^{+0.018}$ & 0.0343 $\pm$ 0.0020  \\
		5-8 & 0.299$_{-0.030}^{+0.030}$ & 0.0613 $\pm$ 0.0026  &  5-8 & 0.243$_{-0.023}^{+0.024}$ & 0.0445 $\pm$ 0.0022  \\
		5-8 & 0.366$_{-0.037}^{+0.038}$ & 0.0822 $\pm$ 0.0031  &  5-8 & 0.295$_{-0.028}^{+0.029}$ & 0.0534 $\pm$ 0.0022  \\
		5-8 & 0.451$_{-0.047}^{+0.046}$ & 0.1121 $\pm$ 0.0046  &  5-8 & 0.359$_{-0.035}^{+0.035}$ & 0.0777 $\pm$ 0.0030  \\
		5-8 & 0.554$_{-0.057}^{+0.061}$ & 0.1548 $\pm$ 0.0064  &  5-8 & 0.437$_{-0.043}^{+0.043}$ & 0.1022 $\pm$ 0.0047  \\
		5-8 & 0.687$_{-0.072}^{+0.079}$ & 0.2171 $\pm$ 0.0085  &  5-8 & 0.531$_{-0.052}^{+0.055}$ & 0.1410 $\pm$ 0.0061  \\
		5-8 & 0.860$_{-0.094}^{+0.101}$ & 0.3089 $\pm$ 0.0112  &  5-8 & 0.650$_{-0.064}^{+0.069}$ & 0.1913 $\pm$ 0.0075  \\
		5-8 & 1.082$_{-0.121}^{+0.136}$ & 0.4432 $\pm$ 0.0165  &  5-8 & 0.800$_{-0.081}^{+0.086}$ & 0.2662 $\pm$ 0.0106  \\
		5-8 & 1.372$_{-0.154}^{+0.192}$ & 0.6393 $\pm$ 0.0227  &  5-8 & 0.982$_{-0.097}^{+0.114}$ & 0.3639 $\pm$ 0.0143  \\
		5-8 & 1.760$_{-0.197}^{+0.282}$ & 0.9261 $\pm$ 0.0343  &      &                         &                      \\
		&                           &                      &      &                         &                      \\
		0-4 & 0.423$_{-0.023}^{+0.022}$ & 0.0985 $\pm$ 0.0038  &  0-4 & 0.421$_{-0.021}^{+0.021}$ & 0.0959 $\pm$ 0.0040  \\
		0-4 & 0.470$_{-0.026}^{+0.025}$ & 0.1145 $\pm$ 0.0049  &  0-4 & 0.466$_{-0.024}^{+0.023}$ & 0.1123 $\pm$ 0.0053  \\
		0-4 & 0.525$_{-0.029}^{+0.027}$ & 0.1370 $\pm$ 0.0046  &  0-4 & 0.516$_{-0.027}^{+0.026}$ & 0.1330 $\pm$ 0.0047  \\
		0-4 & 0.584$_{-0.033}^{+0.032}$ & 0.1617 $\pm$ 0.0054  &  0-4 & 0.572$_{-0.030}^{+0.029}$ & 0.1578 $\pm$ 0.0053  \\
		0-4 & 0.653$_{-0.036}^{+0.037}$ & 0.1929 $\pm$ 0.0061  &  0-4 & 0.633$_{-0.032}^{+0.034}$ & 0.1853 $\pm$ 0.0065  \\
		0-4 & 0.733$_{-0.043}^{+0.041}$ & 0.2312 $\pm$ 0.0072  &  0-4 & 0.704$_{-0.038}^{+0.036}$ & 0.2198 $\pm$ 0.0070  \\
		0-4 & 0.822$_{-0.048}^{+0.048}$ & 0.2777 $\pm$ 0.0081  &  0-4 & 0.782$_{-0.042}^{+0.042}$ & 0.2563 $\pm$ 0.0073  \\
		0-4 & 0.924$_{-0.054}^{+0.056}$ & 0.3320 $\pm$ 0.0084  &  0-4 & 0.870$_{-0.047}^{+0.047}$ & 0.3048 $\pm$ 0.0090  \\
		0-4 & 1.045$_{-0.064}^{+0.064}$ & 0.4002 $\pm$ 0.0105  &  0-4 & 0.969$_{-0.052}^{+0.055}$ & 0.3602 $\pm$ 0.0100  \\
		0-4 & 1.184$_{-0.075}^{+0.074}$ & 0.4894 $\pm$ 0.0121  &  0-4 & 1.084$_{-0.060}^{+0.061}$ & 0.4324 $\pm$ 0.0109  \\
		0-4 & 1.344$_{-0.086}^{+0.089}$ & 0.5974 $\pm$ 0.0145  &  0-4 & 1.214$_{-0.070}^{+0.068}$ & 0.5218 $\pm$ 0.0130  \\
		0-4 & 1.536$_{-0.103}^{+0.104}$ & 0.7381 $\pm$ 0.0186  &  0-4 & 1.357$_{-0.075}^{+0.083}$ & 0.6161 $\pm$ 0.0158  \\
		0-4 & 1.762$_{-0.123}^{+0.124}$ & 0.9148 $\pm$ 0.0213  &  0-4 & 1.527$_{-0.087}^{+0.094}$ & 0.7434 $\pm$ 0.0200  \\
		0-4 & 2.031$_{-0.145}^{+0.152}$ & 1.1349 $\pm$ 0.0258  &  0-4 & 1.721$_{-0.101}^{+0.108}$ & 0.8969 $\pm$ 0.0218  \\
		0-4 & 2.362$_{-0.179}^{+0.182}$ & 1.4270 $\pm$ 0.0298  &  0-4 & 1.943$_{-0.113}^{+0.129}$ & 1.0722 $\pm$ 0.0270  \\
		0-4 & 2.760$_{-0.216}^{+0.228}$ & 1.7806 $\pm$ 0.0360  &  0-4 & 2.203$_{-0.131}^{+0.152}$ & 1.2842 $\pm$ 0.0294  \\
		0-4 & 3.252$_{-0.263}^{+0.293}$ & 2.2670 $\pm$ 0.0415  &  0-4 & 2.500$_{-0.146}^{+0.188}$ & 1.4813 $\pm$ 0.0259  \\
		0-4 & 3.863$_{-0.318}^{+0.391}$ & 2.8449 $\pm$ 0.0448  &      &                           &                      \\
		\hline
	\end{tabular}
	\caption{Relative proton light yield data for EJ-200 and EJ-208. The ID number corresponds to the observation detectors used in the assessment, as outlined in Table~\ref{locationsDec}.}
	\label{resultsTable}
\end{table*}

\begin{table*}
	\centering
	\renewcommand{\arraystretch}{1.2}
	\setlength{\tabcolsep}{4pt}
	\begin{tabular}{ccc|ccc}
		\hline
		\multicolumn{3}{c|}{EJ-204 High Gains} & \multicolumn{3}{c}{EJ-204 Low Gains}   \\
		Observation  & Proton Recoil   & Relative Light Yield  & Observation  & Proton recoil  & Relative Light Yield  \\
		       Detector ID     &  Energy [MeV] &  [dimensionless]   &   Detector ID & energy [MeV] & [dimensionless] \\
		\hline
		9-10 & 0.047$_{-0.007}^{+0.007}$ &  0.0090 $\pm$ 0.0006 &  9-10 & 0.175$_{-0.028}^{+0.026}$ & 0.0309 $\pm$ 0.0020 \\
		9-10 & 0.064$_{-0.010}^{+0.009}$ &  0.0114 $\pm$ 0.0009 &  9-10 & 0.236$_{-0.034}^{+0.041}$ & 0.0479 $\pm$ 0.0028 \\
		9-10 & 0.086$_{-0.013}^{+0.013}$ &  0.0149 $\pm$ 0.0009 &  9-10 & 0.320$_{-0.044}^{+0.061}$ & 0.0614 $\pm$ 0.0051 \\
		9-10 & 0.116$_{-0.017}^{+0.019}$ &  0.0186 $\pm$ 0.0011 &       &                           &                     \\
		9-10 & 0.161$_{-0.026}^{+0.023}$ &  0.0259 $\pm$ 0.0022 &  5-8  & 0.244$_{-0.023}^{+0.024}$ & 0.0451 $\pm$ 0.0021 \\
		9-10 & 0.216$_{-0.032}^{+0.036}$ &  0.0393 $\pm$ 0.0026 &  5-8  & 0.297$_{-0.029}^{+0.029}$ & 0.0588 $\pm$ 0.0028 \\
		9-10 & 0.291$_{-0.040}^{+0.055}$ &  0.0563 $\pm$ 0.0052 &  5-8  & 0.363$_{-0.037}^{+0.036}$ & 0.0797 $\pm$ 0.0034 \\
		&                            &                      &  5-8  & 0.444$_{-0.046}^{+0.045}$ & 0.1079 $\pm$ 0.0047 \\
		5-8 & 0.166$_{-0.016}^{+0.016}$ & 0.0281 $\pm$ 0.0019   &  5-8  & 0.545$_{-0.055}^{+0.058}$ & 0.1473 $\pm$ 0.0064 \\
		5-8 & 0.202$_{-0.020}^{+0.018}$ & 0.0363 $\pm$ 0.0021   &  5-8  & 0.672$_{-0.069}^{+0.075}$ & 0.2053 $\pm$ 0.0085 \\
		5-8 & 0.244$_{-0.024}^{+0.024}$ & 0.0464 $\pm$ 0.0024   &  5-8  & 0.837$_{-0.090}^{+0.095}$ & 0.2894 $\pm$ 0.0115 \\
		5-8 & 0.297$_{-0.029}^{+0.029}$ & 0.0615 $\pm$ 0.0034   &  5-8  & 1.044$_{-0.113}^{+0.128}$ & 0.4105 $\pm$ 0.0162 \\
		5-8 & 0.363$_{-0.037}^{+0.035}$ & 0.0830 $\pm$ 0.0035   &  5-8  & 1.316$_{-0.144}^{+0.175}$ & 0.5885 $\pm$ 0.0223 \\
		5-8 & 0.444$_{-0.045}^{+0.045}$ & 0.1106 $\pm$ 0.0049   &  5-8  & 1.678$_{-0.186}^{+0.248}$ & 0.8470 $\pm$ 0.0346 \\
		5-8 & 0.544$_{-0.056}^{+0.058}$ & 0.1507 $\pm$ 0.0065   &  5-8  & 2.157$_{-0.231}^{+0.376}$ & 1.2266 $\pm$ 0.0508 \\
		5-8 & 0.671$_{-0.069}^{+0.074}$ & 0.2092 $\pm$ 0.0095   &       &                           &                     \\
		5-8 & 0.834$_{-0.090}^{+0.094}$ & 0.2964 $\pm$ 0.0114   &  0-4  & 0.529$_{-0.029}^{+0.027}$ & 0.1395 $\pm$ 0.0044 \\ 
		5-8 & 1.040$_{-0.111}^{+0.127}$ & 0.4183 $\pm$ 0.0169   &  0-4  & 0.588$_{-0.032}^{+0.032}$ & 0.1638 $\pm$ 0.0054 \\ 
		5-8 & 1.308$_{-0.141}^{+0.175}$ & 0.5964 $\pm$ 0.0244   &  0-4  & 0.655$_{-0.036}^{+0.036}$ & 0.1947 $\pm$ 0.0065 \\ 
		5-8 & 1.661$_{-0.179}^{+0.249}$ & 0.8368 $\pm$ 0.0343   &  0-4  & 0.734$_{-0.042}^{+0.039}$ & 0.2323 $\pm$ 0.0070 \\ 
		&                           &                       &  0-4  & 0.819$_{-0.046}^{+0.047}$ & 0.2770 $\pm$ 0.0075 \\ 
		0-4 & 0.528$_{-0.028}^{+0.028}$ & 0.1395 $\pm$ 0.0039   &  0-4  & 0.920$_{-0.054}^{+0.053}$ & 0.3310 $\pm$ 0.0089 \\ 
		0-4 & 0.588$_{-0.032}^{+0.031}$ & 0.1670 $\pm$ 0.0053   &  0-4  & 1.033$_{-0.061}^{+0.062}$ & 0.3986 $\pm$ 0.0104 \\ 
		0-4 & 0.653$_{-0.034}^{+0.037}$ & 0.1984 $\pm$ 0.0052   &  0-4  & 1.167$_{-0.072}^{+0.070}$ & 0.4877 $\pm$ 0.0127 \\ 
		0-4 & 0.732$_{-0.041}^{+0.040}$ & 0.2358 $\pm$ 0.0060   &  0-4  & 1.317$_{-0.079}^{+0.086}$ & 0.5890 $\pm$ 0.0138 \\ 
		0-4 & 0.817$_{-0.045}^{+0.047}$ & 0.2831 $\pm$ 0.0075   &  0-4  & 1.500$_{-0.097}^{+0.097}$ & 0.7242 $\pm$ 0.0170 \\ 
		0-4 & 0.917$_{-0.053}^{+0.053}$ & 0.3402 $\pm$ 0.0083   &  0-4  & 1.711$_{-0.115}^{+0.114}$ & 0.8923 $\pm$ 0.0214 \\ 
		0-4 & 1.031$_{-0.061}^{+0.061}$ & 0.4092 $\pm$ 0.0107   &  0-4  & 1.958$_{-0.133}^{+0.139}$ & 1.1057 $\pm$ 0.0250 \\ 
		0-4 & 1.162$_{-0.070}^{+0.071}$ & 0.4949 $\pm$ 0.0112   &  0-4  & 2.260$_{-0.163}^{+0.164}$ & 1.3796 $\pm$ 0.0308 \\ 
		0-4 & 1.311$_{-0.078}^{+0.086}$ & 0.5980 $\pm$ 0.0138   &  0-4  & 2.620$_{-0.196}^{+0.202}$ & 1.7280 $\pm$ 0.0363 \\ 
		0-4 & 1.492$_{-0.096}^{+0.096}$ & 0.7368 $\pm$ 0.0178   &  0-4  & 3.058$_{-0.235}^{+0.254}$ & 2.1837 $\pm$ 0.0422 \\ 
		0-4 & 1.703$_{-0.114}^{+0.112}$ & 0.9035 $\pm$ 0.0208   &  0-4  & 3.597$_{-0.285}^{+0.327}$ & 2.7768 $\pm$ 0.0550 \\ 
		0-4 & 1.940$_{-0.126}^{+0.143}$ & 1.0978 $\pm$ 0.0229   &  0-4  & 4.272$_{-0.347}^{+0.432}$ & 3.5216 $\pm$ 0.0640 \\ 
		0-4 & 2.222$_{-0.140}^{+0.182}$ & 1.2892 $\pm$ 0.0202   &  0-4  & 5.094$_{-0.389}^{+0.625}$ & 4.3478 $\pm$ 0.0626 \\ 
		\hline
	\end{tabular}
	\caption{Relative proton light yield data for EJ-204. The ID number corresponds to the observation detectors used in the assessment, as outlined in Table~\ref{locationsDec}.}
	\label{resultsTable204}
\end{table*}

\section*{References}

\bibliography{bibfile3}

\end{document}